\documentstyle [aps,epsf]{revtex}

\begin{document}
\twocolumn[\hsize\textwidth\columnwidth\hsize
           \csname @twocolumnfalse\endcsname

\draft
\title{Magnetotransport with two occupied subbands in a Si(100)
inversion layer}
\author{S.\ G.\ Semenchinsky\cite{AA}, L.\ Smr\v{c}ka,
and J.\ Stehno}
\address{ Institute of Physics, Academy of Science of the Czech
Republic,\\ Cukrovarnick\'{a} 10, 162 00 Prague 6, Czech Republic}

\date{Received ......}
\maketitle
\begin{abstract}
We have studied an electron transport in inversion layers of
high-mobility Si(100) samples. At high electron concentrations
and temperatures below 4.2 K, two series of Shubnikov-de Haas
oscillations have been observed. The temperature damping of the
second series oscillations indicates that the second occupied
subband belongs to the first energy level of the
fourfold-degenerate ladder $0'$. Samples with two occupied
subbans exhibit a strong anomalous negative magnetoresitance,
reaching $\rm \approx 25 \%$ of a zero field value at $B =$ 12 T.
The resistance decrease is more pronounced for lower temperatures
and higher electron concentrations. We explain this behaviour by
an increase of the second subband mobility due to the
freezing-out of the scattering of $0'$ electrons. Based on the
measured periods of SdH oscillations, we conclude that the
electrons are distributed inhomogeneously beneath the sample
gate.
\end{abstract}

\vskip2pc]

\narrowtext
The $n$-type inversion layer on a (100) surface of $p$-type Si
has been the subject of investigation for many years (for review
see \cite{ando}). The silicon is an indirect gap semiconductor
having six minima (valleys) in the conduction band along the
(100) direction. In the case of (100)-oriented wafers, two of the
valleys in the Brillouin zone are oriented perpendicular to the
$\rm Si/SiO_2$ interface (with $m_z = 0.916$), while the other
four are parallel and have $m_z = 0.19$. A quantum well is formed
at the interface by applying a positive voltage $V_g$ to the
gate, and the electron motion in $z$ direction is quantized. The
difference in $m_z$ causes a splitting of the energy spectrum
into two independent ladders of levels: a twofold-degenerate
ladder denoted $0$, and a fourfold-degenerate ladder $0'$. Due to
the higher effective mass in the $z$ direction, the lowest energy
level in the potential well belongs to the $0$-ladder.

Disagreement exists in the literature on the electron
concentration at which the Fermi energy crosses the next energy
level and on the question whether it forms the bottom of the
second subband of the \lq unprimed\rq{} ladder, or the first
subband of the \lq primed\rq{} ladder \cite{tsui,how}. More
recently this problem was discussed in \cite{klap}, based on the
combination of measurements of the gate voltage dependence of the
magnetoresistance and the Hall voltage. It was shown that the
mobility of the second subband is higher than in the first
subband (but relatively low), and that it linearly decreases with
the temperature. This temperature dependence was tentatively
explained by the temperature-dependent screening of the elastic
scattering in the second subband. It was concluded from the
experimental data that the second subband is the lowest subband
of the fourfold-degenerate ladder, but the corresponding
Shubnikov-de Haas (SdH) oscillations have never been observed
explicitly. Here we report, to our knowledge for the first time,
the magnetoresistance measurements on high-mobility Si(100)
samples, showing clearly both series of Shubnikov-de Haas
oscillations and supporting the idea that the second occupied
subband belongs to the first energy level of the
fourfold-degenerate ladder.

We employed in our experiments five Hall bar samples prepared
from three different wafers. The wafers were made from $p$-type
silicon with resistivity $\rm 20\Omega cm$ at the room
temperature, their crystallografic orientation was checked by
X-ray diffraction and found (100) with an accuracy within
$\rm\pm 0.5\%$. The edges of samples, defined by the gate
geometry, were oriented along (110) direction. Results of
measurements were very similar for all samples used. Therefore,
we present data obtained for two of them, denoted as \lq sample
1\rq{} and \lq sample 2\rq{}. The maximum mobility of electrons
in 2D inversion layers was about 2.7 $\rm m^2/Vs$ for the sample
1, and 2.0 $\rm m^2/Vs$ for the sample 2 at the temperature $T
= $ 4.2 K. The dimensions of samples were $\rm 5\times 0.8\,
mm^2$, with the distance between potential leads $\rm 2.5\, mm$,
(sample 1) and $\rm 2.5\times 0.25\, mm$, with the distance
between leads $\rm 0.625\, mm$ (sample 2). The 0.2 $\rm \mu m$
thick $\rm SiO_2$ gate insulator allowed to feed the electrons
into an inversion layer with the maximum concentration $N_s
\approx \rm 1.3\times 10^{17} m^{-2}$. There is a linear
dependence between $N_s$ and $V_g$ determined by the sample
capacitance; the rate of filling, $N_s/V_g$, was approximately
$\rm 1.1\times 10^{15} m^{-2}V^{-1}$ for both samples. Shubnikov
de Haas oscillations were investigated either as a function of
$V_g$, for a constant magnetic field $B$, or as a function of
$B$, assuming fixed $V_g$. We used a DC measuring method, the
current through the sample was varied in the range 0.1-100$\rm
\mu A$, and the voltage between the potential leads was measured
by a digital voltmeter. The samples, immersed in the liquid $\rm
^4He$, were cooled down to 1.9 K in magnetic fields up to 12 T.

\begin{figure}
\begin{center}
\leavevmode
\epsfbox{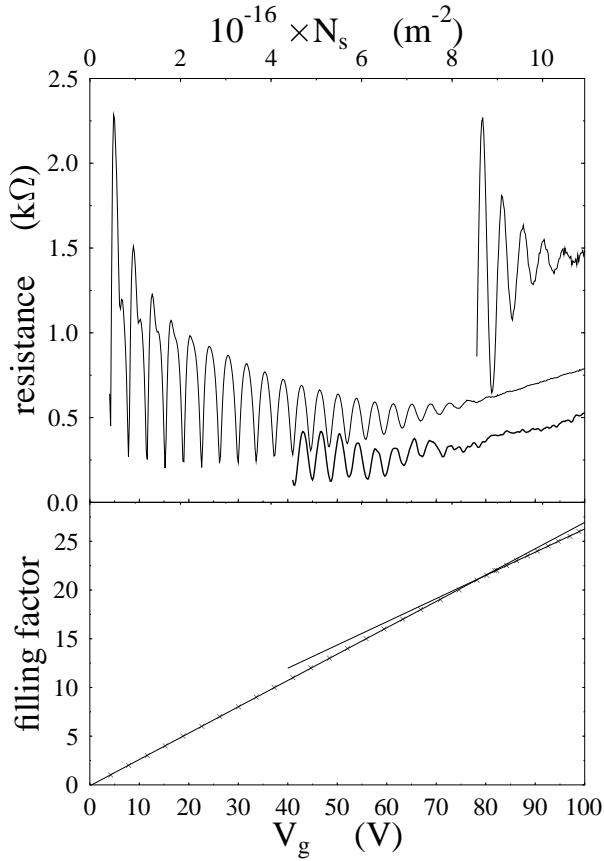}
\end{center}
\caption{Shubnikov - de Haas oscillations and filling factors for
sample 2 at $B$ = 4.18 T, $I$ = 10 $\mu$A, and $T$ = 1.9 K. The
upper right curve is a derivative of a main one with respect to
$V_g$, with an arbitrary $y$-scale (thin lines). The thick line
shows results of a measurement using a reduced current 0.1
$\mu$A.}
\label{f1}
\end{figure}

Figure \ref{f1} presents the selected measurements of
magnetoresistance as a function of the gate voltage $V_g$.

Firstly, the upper panel shows the \lq standard\rq{} data
obtained using the measuring current $I =10 \mu$A; these results
agree quite well with results of previous measurements
\cite{klap}. The observed oscillations of magnetoresistance
correspond to filling of Landau levels of the first subband and
they are periodically dependent on $V_g$ for the electron
concentration lower than $\approx \rm 8 \times 10^{16} m^{-2}$.
Due to the spin degeneracy, and the twofold valley degeneracy,
there are $4|e|B/h$ electrons per unit area in each Landau level.
The amplitude of oscillations drops down rapidly for higher
electron concentrations, and their period decreases. To help
resolve the oscillations, the inset presents the derivative of
the magnetoresistance curve with respect to $V_g$ for higher
$N_s$ region.
The lower panel shows the filling factor of
electrons in the first subband, determined from the minima of the
SdH oscillations shown above. For $V_g < \rm 80\, V$, the
concentration $N_{s,1}$ follows the straight line, implying
$N_{s,1} = N_s$. For higher $V_g$, a single period oscillations
remain present, but the electron concentration in the first
subband is lower, since a part of electrons is fed into the
higher subband.

Secondly, a part of an \lq anomalous\rq{} magnetoresistance
curve is presented in the upper panel of figure \ref{f1}. The
measuring procedure was the same as described above, but the
current through the sample, $I$, was reduced one hundred times,
to 0.1 $\mu$A. In that case the second series oscillations became
visible and the decrease of the magnetoresistance is observed. It
is important to stress that the resistance measured
at $B = 0$ was identical for both measuring currents, i.e. the
same curves were obtained for $I =$ 10 $\mu$A and $I =$ 0.1
$\mu$A. Therefore, it is not possible to explain the drop of
magnetoresistance simply as due to the reduced heating of 2D
electron gas. If it would be the case, the reduced electron
temperature should imply e.g an increase of amplitude of SdH
oscillations of the first subband electrons. Instead of it, the
decrease of amplitudes was observed, as if a part of a current,
originally carried by the first subband electrons, was taken over
by the second subband electrons with higher mobility. This
indicates that only the electrons from the second subband are
influenced by the the current magnitude.

\begin{figure}
\begin{center}
\leavevmode
\epsfbox{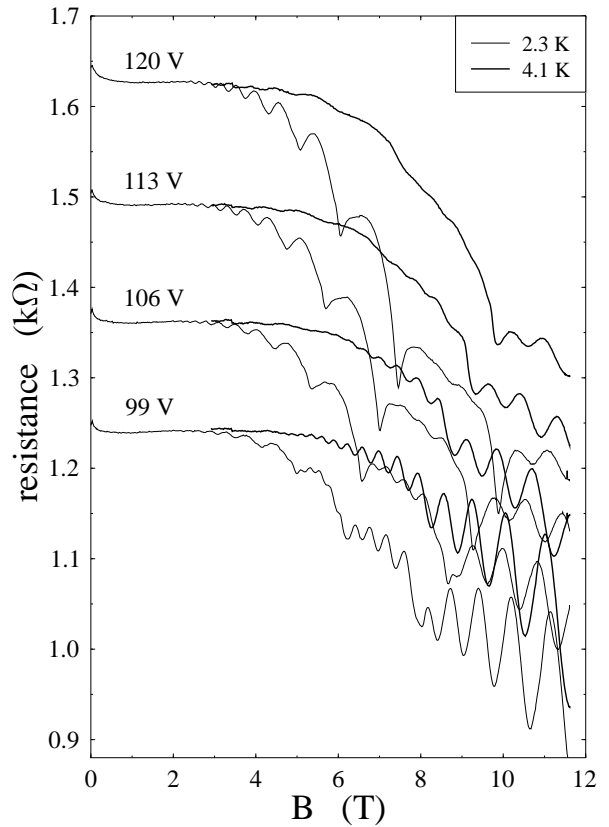}
\end{center}
\caption{ Magnetoresistance oscillations for several electron
concentrations   and two temperatures. Sample 1, $I$ = 5 $\mu$A.}
\label{f2}
\end{figure}
These unusual results motivated us to more laborious
investigation of the field dependence of magnetoresistance with
the fixed gate voltage $V_g$ and the variable magnetic field. In
that case the coexistence of two series of oscillations can be
studied, together with the development of the negative
magnetoresistance. Examples of
such curves, obtained for four gate voltages at two different
temperatures, and employing a relatively high current $I = \rm
5\, \mu A$, are presented in figure \ref{f2}. In the lowest magnetic
fields, the negative magnetoresistance, attributed to the
localization, is observed for all four $V_g$, followed by
a slight increase, which lasts till $B \approx 3$ T. In this low
field region, the magnetoresistance practically does not depend
on the sample temperature. For higher fields, the SdH
oscillations appear, corresponding to both subbands, and their
emergence is accompanied by onset of a strong negative
magnetoresistance. The field dependence of smooth parts of curves
is very sensitive to the temperature, for lower temperatures
a more pronounced resistance decrease occurs, similarly as in the
case of reduced current density, as presented in figure \ref{f1}.

\begin{figure}
\begin{center}
\leavevmode
\epsfbox{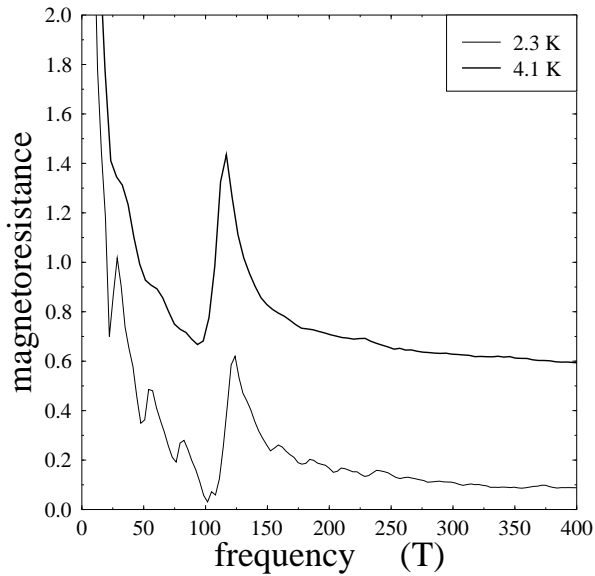}
\end{center}
\caption{Absolute values of Fourier transforms of high-field
parts of magnetoresistance curves measured at $V_g =$ 99 V
and shown in figure \protect \ref{f2}.}
\label{f3}
\end{figure}

The magnetoresistance curves, measured as a function of $B$ at
various temperatures, allow us to derive important
characteristics of the second subband. Firstly, we have
determine the cyclotron effective mass from the temperature
damping of the oscillations amplitudes, and found $m_c = \rm
0.43 \pm 0.03$, in good agreement with a theoretical value, $m_c
= \rm 0.417$. This implies that the first subband of the $0'$
ladder is occupied, in agreement with conclusions of references
\cite{tsui} and \cite{klap}.

Both series of oscillations are periodical functions of the
inverse magnetic field and their frequencies are linear
functions of concentrations of carriers in the individual
subbands. Examples of Fourier transforms of two curves, measured
at $V_g =$ 99 V and presented in figure \ref{f2}, are shown in
figure \ref{f3}. The lines are offset for clarity. Peaks
corresponding to two different frequencies are clearly seen. The
first subband oscillations are almost sinusoidal and, therefore,
described by a single peak which has somewhat lower amplitude for
the curve measured at lower temperature. The second subband
oscillations contains a large number of higher harmonics and the
corresponding peaks are more visible at lower temperature curve.
This can be understood having in mind a semi-elliptic form of
oscillations. The smooth
background is due to the smooth part of the negative
magnetoresistance which was not removed before transforming the
data.
\begin{figure}
\begin{center}
\leavevmode
\epsfbox{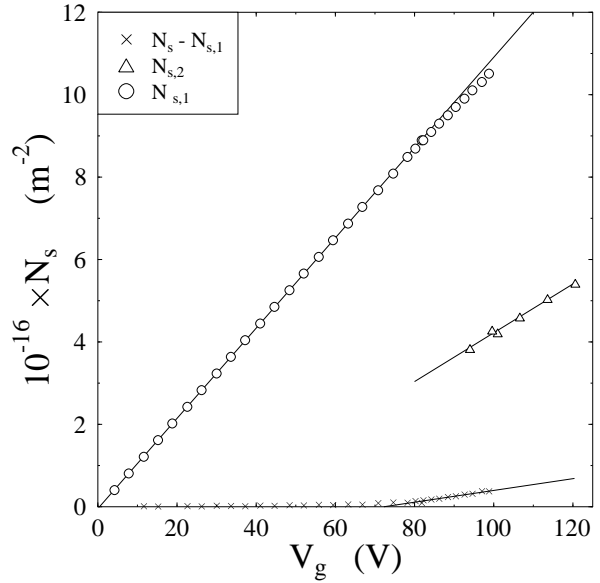}
\end{center}
\caption{Electron concentrations calculated
from periods of magnetoresistance oscillations.
Also a difference between the full concentration $N_s$, and the first
subband electrons concentration $N_{s,1}$ is shown. Sample 1, $I$ = 0.1
$\mu$A.}
\label{f4}
\end{figure}

Both $N_{s,1}$ and $N_{s,2}$ determined from the periods
of SdH oscillations are shown in figure \ref{f4}.
There are only a few points of $N_{s,2}(V_g)$ dependence and,
therefore, the linear extrapolation to lover gate voltage has
only a limited meaning. Nevertheless, it indicates a lower
threshold of occupation of the second subband than reported
previously. The results also show that the ratio of two periods
of oscillations is approximately 4.5 for all four investigated
gate voltages, which means $N_{s,1}/N_{s,2} \approx $ 2.25, if
the fourfold degeneracy of the second subband is assumed.
Traditionally, \cite{tsui,how,klap}, the concentration of
electrons in the second subband, $N_{s,2}$, is deduced from the
decrease of the period of the first subband oscillations at
higher gate voltages, such as shown in figure \ref{f1}. If the
homogeneous distribution of electrons beneath the sample gate is
assumed, $N_{s,2}$ is given by the difference $N_{s}
- N_{s,1}$. Our direct determination of $N_{s,1}$ and $N_{s,2}$
from the oscillation periods does not confirm results of this
approach in the case of $N_{s,2}$.

More series of SdH oscillations were already reported for
inversion layers on surfaces slightly tilted from (100)
\cite{math} and for layers on (100) surfaces subject to either
uniaxial \cite{abst} or biaxial stress \cite{fang}. In the first
case, only the first subband electrons are involved and the
reason for additional periods is lifting of the twofold valley
degeneracy caused by the tilting and the inter-valley
interaction. In the latter case, the crystal symmetry is lowered
due to the stress and, consequently, the energy of $0'$ levels
shifted below the Fermi energy level.

We do not think that our second series oscillations can be
satisfactory explained by the above mentioned mechanisms. The
surface orientation is (100) with high accuracy and the only
sources of stress are the different thermal expansion
coefficients of Si, $\rm SiO_2$ and the gate material. On the
other hand, we would like to turn attention to another
possibility, not mentioned previously, why the bottom of the
first subband of the $0'$ ladder can be shifted below the Fermi
energy only in a part of the sample area: the inhomogeneity of
distribution of electrons beneath the sample gate.

In our samples, the area occupied by a two-dimensional electron
gas is defined by the gate geometry rather then by fixed ionized
donors, as it is the case in modulation doped heterostructures.
The concentration of electrons $N_s$ is related to $V_g$ by $N_s
= \epsilon V_g/|e|d$ where $\epsilon$ is a dielectric constant of
the oxide and $d$ is its thickness. The charged gate and the 2D
electron system act as a parallel plate capacitor; the plates
containing the free charge keep the constant potentials through
their areas, rather then the constant concentration of carriers.
Therefore, the free charge density can fluctuate inside the sample,
e.g. due to the surface roughness and/or an inhomogeneous
distribution of charged impurities in the depletion layer.
Moreover, the free charge should accumulate near the plate edges,
to compensate the Coulomb repulsion of carriers from the sample
interior. The electrostatic corrections to the finite dimension
of a capacitor \cite{max} predict that the extra charge is
localized in narrow strips near the edges, with the width
proportional to $d$, and that  the excess concentration,
$\Delta N_s(x)$, diverges when $x$, the distance measured from
the edge, approaches zero. In real life the concentration will
stay finite, nevertheless we can assume that it first rises near
an edge, before it vanishes. Thus, we consider that the narrow
channels near the edges concentrate a part of electrons from the
second subband and contribute substantially to the sample
conductivity.

The high sensitivity of the observed results to the measuring
current density and to the temperature suggests that the
electron-electron interaction can be involved. Firstly, the
temperature-dependent screening of the second subband electron
scattering can play
a role, as described in \cite{klap}. Secondly, the properties of
the second subband electrons should be strongly influenced by
quantization into the Landau levels
for $\hbar\omega > k_BT$. Note that the onset of the
negative-dependent magnetoresistance occurs just when the above
condition is satisfied.

 The Landau levels of a single subband never cross and their
shape more or less copies a shape of a smooth confining potential
and exhibits a characteristic step-like structure near abrupt
potential steps. The Landau levels of electrons from two
different subbands, 0 and 0', can cross, as they will see the
shape of a sample potential in a different way. Firstly, due
to a large difference between the number of Landau levels near
the Fermi energy and, secondly, due to the difference in the
effective masses. The points of crossing can represent the \lq
hot\rq{} points of an intensive intersubband scattering which can
substantially contribute to the observed effects and can be
suppressed at low temperatures and high magnetic fields.

The increase of the mobility of the second subband electrons causes
the redistribution of the current density and more current is
conducted by them. As a consequence, the amplitudes of second
series oscillations growth to detriment of amplitudes of the
first series oscillations. The extension of present measurements
down to the milikelvin temperatures region, which is now in
progress, can probably shed more light to the nature of this new
interesting effects.

We thank to V.\ Borzenets, L.\ Skrbek and J.\ \v{S}ebek for
stimulating discussions and their help with cryogenic
measurements. We are also grateful to Z.\ V\'{y}born\'{y} for his
help with sample preparation. This work was in part supported by
International
Science Foundation through the Grant RHD 000, by the Grant Agency
of the Czech Republic under Grant No 202/93/0027 and by the Academy
of Science of the Czech Republic under Contract No. 110 423.

\end{document}